\documentclass[twocolumn,nofootinbib,aps,prl,showpacs]{revtex4-1}
\usepackage{graphicx}
\usepackage{float}
\usepackage{color}
\usepackage{graphicx}
\usepackage{dcolumn}
\usepackage{amsmath}
\usepackage{amsfonts}
\usepackage{hyperref}
\usepackage{epstopdf}
\usepackage{color}
\newcommand{\mathsym}[1]{{}}

\begin{document}
\title{Jets or vortices - what flows are generated by an inverse turbulent cascade?
}
\author{Anna Frishman$^{1,2}$, Jason Laurie$^{1,3}$ and Gregory Falkovich$^{1,4}$}
\affiliation{$^{1}$Department of Physics of Complex Systems, Weizmann Institute of Science, Rehovot 76100 Israel\\
$^{2}$Princeton Center for Theoretical Science, Princeton University, Princeton, New Jersey 08544, USA\\
$^{3}$Mathematics Group, School of Engineering and Applied Science, Aston University, Birmingham, B4 7ET, UK\\
$^{4}$Institute for Information Transmission Problems, Moscow, 127994 Russia}
\date{\today}
\begin{abstract}
An inverse cascade -
 energy transfer to
progressively larger scales - is a salient feature of two-dimensional turbulence. If the cascade reaches the system scale, it creates a coherent flow  expected to have the largest available scale and conform with the symmetries of the domain. In a doubly periodic rectangle,  the mean flow with zero total momentum was therefore believed to be unidirectional, with two jets along the short side;
while for an aspect ratio close to unity, a vortex dipole was expected. Using direct numerical simulations, we show that in fact neither the box symmetry is respected nor the largest scale is realized: the flow is never purely unidirectional since the inverse cascade
produces coherent vortices, whose number and relative motion are determined by the aspect ratio. This spontaneous symmetry breaking is closely related to the hierarchy of averaging times. Long-time averaging restores translational invariance due to vortex wandering along one direction, and gives jets whose profile, however, can be deduced neither from the largest-available-scale argument, nor from the often employed maximum-entropy principle or quasi-linear approximation.

\end{abstract}


\maketitle

\noindent{\it \textbf{Introduction.}}
An inverse cascade is a counter-intuitive process of self-organization of two-dimensional turbulence. In an infinite medium, the cascade creates vortices (vortex rings in curved space~\cite{falkovich_turbulence_2014}) of ever-increasing size, while in a finite domain it eventually 
forms a flow coherent in the entire system.
That flow is expected to become  universal, i.e. independent of  the forcing, when forcing scale goes to zero, keeping the inverse energy flux finite.
Predicting the form of the flow in various settings is one of the central problems of turbulence theory.
There are three known ways to address this issue. The first way is qualitative: to look for the flow with the largest available scale. One quantitative way,  justifiable only in equilibrium,
looks for a flow profile that maximizes
entropy~\cite{miller_statistical_1990,robert_statistical_1991}. The third way is to assume turbulence weak relative to the mean flow and employ a quasi-linear approximation, writing equations for the two-point correlation functions (of velocity or vorticity) to form a closed system~\cite{farrell_structure_2007,marston_statistics_2008,bouchet_stochastic_2014} or using a single-point reduced description of spatial fluxes (of energy, momentum, enstrophy)
~\cite{laurie_universal_2014,falkovich_interaction_2016}.

Perhaps the simplest setting is a rectangle
with periodic boundary conditions (a torus). The system is translation invariant along $x$ and $y$; any nonuniform mean flow breaks one of these symmetries or both.
Flow on a torus may have either contractible streamlines corresponding to vortices or  non-contractible streamlines
corresponding to jets. Jets going around one side may be expected in a (non-square) rectangle where there is no symmetry between directions.
Indeed, the maximal-entropy argument predicts 
a jet directed along the shorter side
(two opposite jets for zero total momentum)~\cite{miller_statistical_1990,robert_statistical_1991,bouchet_statistical_2012}. Furthermore, numerical evidence in non-equilibrium regimes shows two jets 
for a rectangle, and a vortex dipole flow 
for domains close to a square~\cite{bouchet_random_2009}.

One may also try to explain
the appearance of the two flow types via the largest-scale argument: for  an aspect ratio $l_x$ substantially different from one, the largest mode is two opposite jets along the short side. On the other hand, in a square box, the jets can be directed along either side and one may expect a superposition of two sets of jets, which would look like a vortex dipole~\cite{bouchet_statistical_2012}. In this picture, jets are fundamental objects on a torus while the vortex dipole appears only near a degeneracy, when $l_x\approx1$.

In fact, the vortex, created by an extended inverse cascade in a square box of size $L\times L$, cannot be represented as a superposition of jets~\cite{chertkov_dynamics_2007,laurie_universal_2014}, since  the analysis of spatial fluxes of energy and momentum requires the velocity profile $U(r)\propto Const.$ at $r\ll L$~\cite{laurie_universal_2014}; that corresponds to the streamfunction $\psi(r)\propto r$ which cannot be represented as a superposition of orthogonal jets, $\psi(x,y)=\phi(x)+\phi(y)$.
While no details of the forcing spectra were given in~\cite{bouchet_random_2009}, where a dipole-jet transition was observed upon the change of the aspect ratio, it is likely that the forcing scale was comparable to the box size and no pronounced inverse cascade was present.

Here we consider the universal limit of 
the culmination of an extended inverse cascade in a rectangle with low friction.
Numerical modeling reveals how all the expectations are defied by Nature: there is no dichotomy between vortices and jets, which coexist for any aspect ratio.
For domains with a moderate $l_x$, we find two jets and a vortex dipole. Decreasing $l_x$ causes the vortices to wander relative to each other along the jets. The asymptotic state is sensitive to the value of very small uniform friction: as it decreases for a moderate $l_x$, an additional vortex, that can be of either sign, emerges.  When averaged over long times, vortices are smeared into strips, resulting in a two-jet mean flow.
Decreasing $l_x$ causes appearance of additional vortices, and what is more interesting, additional jets,  which persist under long-time averaging.  We thus find that the first two approaches (largest-scale and maximal-entropy) give wrong predictions. The third approach, the quasi-linear approximation, relies on the expectation that the mean flow will be dominated by turbulence in the universal limit. That approximation requires a proper account of the flow geometry and averaging time; it describes correctly the interior of a circular vortex~\cite{laurie_universal_2014}, but fails to describe the global mean flow in a rectangle: the zonal or long time average treats vortices as fluctuations making the latter strong, as our numerics show. Our work thus demonstrates that the principles of organization of an inverse cascade into a mean flow are currently lacking.

{\it \textbf{Jets and vortices}.}
We consider an incompressible flow, $\nabla \cdot \mathbf{v}  =0$, described by the two-dimensional Navier-Stokes equations for a fluid with unit density:
\begin{align}
\partial_t\mathbf{v}+\left(\mathbf{v} \cdot \nabla\right) \mathbf{v}&=-\nabla p-\alpha \mathbf{v}+\nu \nabla^2 \mathbf{v}+ \mathbf{f},
\label{eq:NavStokes}
\end{align}
The force $\mathbf{f}$ acts in a narrow band of scales $l_f\ll L$. The energy injection rate is  $\epsilon = \langle \mathbf{f}\cdot \mathbf{v}\rangle$. We assume that system-size eddies produced by an inverse cascade have turnover times  much shorter than the time of frictional dissipation: $\delta\equiv\epsilon^{-1/3}L^{2/3}\alpha\ll 1$. Then the inverse cascade fed by our small-scale forcing
reaches the system scale producing energy accumulation and mean flow generation. In the steady state at high Reynolds number, $Re=\epsilon^{1/3}l_f^{4/3}/\nu\gg1$, most of the energy is  dissipated by the friction of the mean flow, giving the mean velocity estimate $U\simeq \sqrt{\epsilon/\alpha}$, and the corresponding
turnover time $\tau_m=\sqrt{\alpha L^2/\epsilon}$.

We numerically solve~\eqref{eq:NavStokes} in the vorticity formulation in a periodic rectangle of size $2\pi l_x\times 2\pi$ using  the grid $512l_x \times 512$ with spacing uniform in both directions. We implement a pseudo-spectral method using the $3/2$-dealiasing rule and time step using a forth-order exponential time-difference algorithm. Each Fourier mode of the forcing is  a complex Gaussian random variable, delta-correlated in time, with fixed amplitude equal to $0.1$ in an annulus of width $99 \leq k < 101$. The force scale is defined as $l_f=2\pi/k_f$ with $k_f\approx 100$. In order to provide as large as possible inertial range for the inverse cascade, we use 
hyper-viscosity  $-\nu(-\nabla^2)^{p} \mathbf{v}$ with $p=8$ and $\nu=1\times 10^{-36}$.
Each simulation is run until reaching a statistical steady state verified by monitoring the total energy. We compute the energy dissipation rate via friction using it as
as a measure of the inverse energy flux
and an estimate for
$\epsilon$. All of our data analysis is performed in the statistical steady state.

\begin{figure}[!ht]
\begin{center}
\includegraphics[width=1\columnwidth]{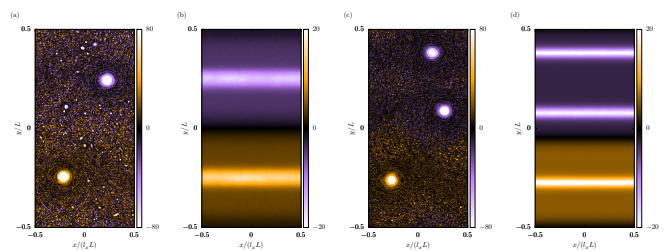}
\caption{(color online) Scaled total vorticity $\sqrt{\alpha L^2/\epsilon}\left(\nabla \times {\bf v}\right)$ for runs A and F respectively: (a) and (c) a snapshot, (b) and (d) a temporal average over time $ 1/\alpha$ with frames shifted to align the velocity maximum  with $y=0$. \label{fig:1}}
\end{center}
\end{figure}

We begin from $l_x < 1$ to see if the emergent mean flow has two opposite jets parallel to $\hat{x}$, with all averaged quantities independent of $x$. We  perform three simulations, denoted by A-C, with $\alpha=1\times 10^{-4}$ and  with different aspect ratios:  $l_x=1/2$, $\delta = 5.58\times 10^{-3}$ (A), $l_x=3/4$, $\delta = 4.86\times 10^{-3}$ (B) and $l_x=1$, $\delta = 4.39\times 10^{-3} $ (C).
The typical vorticity snapshot in a  steady state reveals a surprising feature:  large-scale coherent vortices in addition to jets, see Fig.~\ref{fig:1}(a).
Considering the dynamical generation of the mean flow, the presence of vortices is natural: locally the inverse cascade tends to create vortices, and the anisotropy of the box is felt only when their size is comparable to $l_xL$. Once established, the opposite-signed vortices feed on the constantly created smaller vortices, counteracting the effect of dissipation.

\begin{figure}[!ht]
\begin{center}
\includegraphics[width=1\columnwidth]{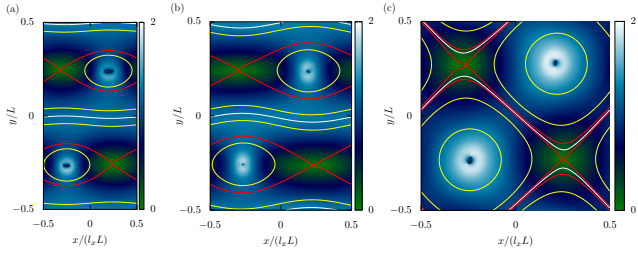}
\caption{(color online) Heat maps of the scaled speed ($\sqrt{\alpha/\epsilon} |\bf{v}|$) averaged over time $\tau_m$: (a) run A, (b) run B, and (c) run C. Overlaid are streamlines, red lines are separatrices, white lines correspond to $\psi=0$.  \label{fig:2}}
\end{center}
\end{figure}

Any meaningful discussion of the emerging mean flow and symmetries must address the averaging times.
For $t \lesssim \tau_m$ the centres of the vortices are effectively pinned. Averaging on such timescales, the mean flow can be characterized by streamlines as presented in Fig.~\ref{fig:2}.
Topologically, the mean flow consists of two distinct regions of contractible streamlines surrounding the centres of the two vortices. In between the two regions, a separatrix should be present.
The positive vortex is at $(-  l_xL/4,-L/4)$ in Fig.~\ref{fig:2}.
The symmetry of the streamline pattern around the vortex center dictates that the separatrix would pass through $(l_xL/4,-L/4)$, which is a stagnation point due to periodicity.
For a square box, the vortices are arranged in a diagonal lattice, with the second vortex located at $(l_xL/4,L/4)$ and the separatrix passing through the stagnation point $( -l_xL/4,L/4)$. A separatrix composed of two straight streamlines connecting the two stagnation points preserves the $x-y$ symmetry, and is therefore expected for a square box where $l_x=1$.
For $l_x<1$, this symmetry is absent and there is no reason for the two stagnation points to lie on the same streamline. We thus expect the separatrix to split into two, giving rise to two regions of non-contractible streamlines, i.e. jets. We indeed observe this splitting for aspect ratios $1/2$ and $3/4$, as seen in Fig.~\ref{fig:2}.
In the square box, we also find the $x-y$ symmetry spontaneously broken by a tiny splitting of the separatrix, creating an opening for weak jets.

\begin{figure}[!ht]
\begin{center}
\includegraphics[width=1\columnwidth]{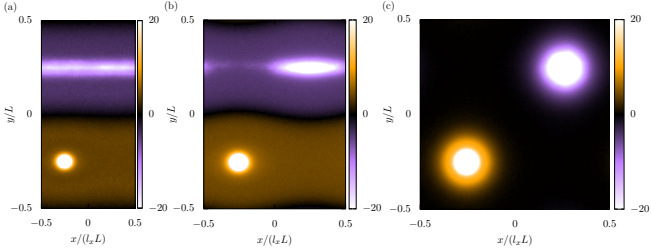}
\caption{(color online) Scaled total vorticity $\sqrt{\alpha L^2/\epsilon}\left(\nabla \times {\bf v}\right)$ averaged over time $  1/\alpha$ in the reference frame of the positive vortex: (a) run A, (b) run B, and (c) run C. \label{fig:3}}
\end{center}
\end{figure}


\begin{figure}[!ht]
\begin{center}
\includegraphics[width=\columnwidth]{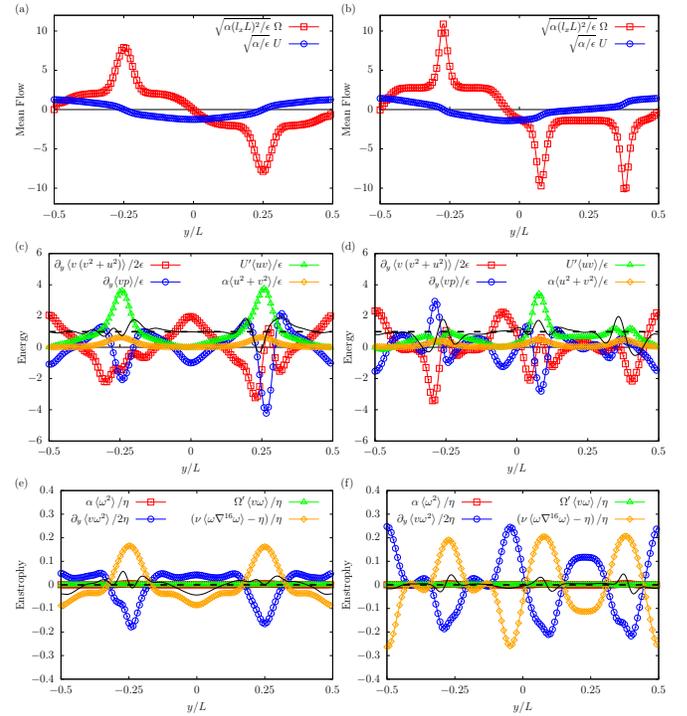}
\caption{(color online) Mean profiles of vorticity and velocity (top), balances of energy (middle) and enstrophy (bottom) for run A (right) and run F (left). Horizontal dashed lines indicate the expected  balance, solid lines indicate the numerical sum. Data were averaged over time exceeding $\tau_d$  with frames shifted to align the velocity maximum with $y=0$. \label{fig:6}}
\end{center}
\end{figure}

On larger timescales, associated to the fluctuations $t \gtrsim \tau_{f}\equiv \epsilon^{-1/3}L^{2/3} $, a collective motion of the vortices, and a relative horizontal motion for $l_x<1$, becomes appreciable, see Fig.~\ref{fig:3} and Fig.~S4 in the supplementary. In particular, for $l_x=1/2$ and $l_x=3/4$,  the diagonal lattice is only one among a continuum of possible configurations. We conjecture that the available configurations are determined by the presence of a minimal distance between the vortex centres.  In a square box, the vortices are almost completely restricted to the diagonal, separated by the maximal possible distance, which suggests that $l_x L/\sqrt{2}$ may be the  minimal vortex separation for arbitrary $l_x$.
Correspondingly, we find that the extent of the horizontal relative motion of the vortices for $l_x=3/4$ is close to $(l_x-\sqrt{2 l_x^2-1})L$, see Fig.~\ref{fig:3}(b) and also Fig.~S1 in the supplementary. This is in agreement with the absence of configurations with inter-vortex separations smaller than $l_x L/\sqrt{2}$, assuming that the vertical relative motion is negligible. For aspect ratios $l_x<1/\sqrt{2}$, the full extent of the vortex relative horizontal motion should be allowed. Indeed, as observed Fig.~\ref{fig:3}(a) for $l_x=1/2 < 1/\sqrt{2}$ the second vortex explores the line $y=L/4$ with equal probability.

For $l_x=1/2$, we did runs for different $\delta$:  $\delta=1.12\times 10^{-2}$ (D), $\delta=2.78\times 10^{-3}$ (E) and $\delta=1.67\times 10^{-3}$ (F).
Astonishingly, the vortex dipole of simulations A and D is replaced by a three-vortex configuration for the low-friction simulations E and F, see Fig.~\ref{fig:1}(c).
In this configuration, the two same-sign vortices lie inside different jets, moving horizontally in opposite directions already on timescales of order $\tau_m$. The third vortex remains on the zero-velocity line between two jets. The distance between the opposite-sign vortices always exceeds the suggested minimal distance, $l_x L/\sqrt{2}$.
In ~\cite{kolokolov_structure_2016}, it was argued that the size of the vortices should grow with decreasing $\delta$ for fixed $l_f$ and $\epsilon$, this contradicts the appearance of additional vortices found here.

How many vortices does the asymptotic $\delta\to 0$ state contain for a given  $l_x$? If indeed there exists a minimal sustainable separation between opposite-sign vortices,
then their number is limited by it.
If adjacent same signed vortices can appear only inside different jets, then at most four vortices, arranged in a diagonal lattice, can be present for $l_x=1/2$. The emergence of such a constricted arrangement out of the three-vortex configuration seems improbable. Thus, three vortices may be the asymptotic state for $l_x=1/2$.

Simulations G and H were done respectively for   $l_x=1/3$, $\delta=6.22\times 10^{-3}$  and $l_x=1/4$, $\delta=6.83\times 10^{-3}$.
With decreasing aspect ratio, not only does the number of vortices increase but also the number of jets: four jets are present for  $l_x=1/4$. This implies that the length of the short side of the box plays a crucial role in determining the jets,
in contradiction to the largest-mode argument. Snapshots of the vorticity field are presented in Fig.~S6 of the supplementary.

{\it \textbf{Long time average: mean flow and fluctuations.}}
In the limit $\delta \to 0$, $\tau_m$ and $\tau_{d}\equiv 1/\alpha$ become well separated and one can average over a time between them. %
For $l_x<1$, while there is random motion along $x$ on such timescales, almost no vertical (collective or relative)  motion of the vortices is observable even for the longest times of order $\tau_d$, on which the square-box mean flow is close to zero, see Fig~S5 in the supplementary.
Thus,  averaging over $t\gg \tau_f\gg\tau_m$ smears vortices into stripes resulting in an effective jet-like mean flow homogeneous in $x$, as is shown in Fig.~\ref{fig:1}(b) and Fig.~\ref{fig:1}(d) for $l_x=1/2$. In Figs.~\ref{fig:1}(b) and~\ref{fig:1}(d) we align frames by the line of maximum (positive) $x$-averaged velocity.
While the size of the vortices only slightly decreases with $\delta$, see Fig.~\ref{fig:1}, the averaged vorticity strip is narrower for smaller $\delta$  due to the suppression of the fluctuations which cause the vertical drift of the vortices.

The statistical homogeneity along $x$  at long times allows for an analysis of the energy and momentum balance similar to that in~\cite{laurie_universal_2014}.
Accordingly, we decompose the velocity into its mean $U(y)$ and fluctuating components: $\mathbf{v}=(u+U,v)$ where  $u\parallel x$, $v\parallel y$ and $\left\langle u\right\rangle= \left\langle v\right\rangle=0$, the average being over time. In the simulations a zonal average along $x$  is added.
We write the steady state conservation of the $x$-momentum and energy neglecting the viscous terms,
assuming  $\nu/\alpha L^2 \ll 1$ and
$Re\to \infty$. Denoting $y$-derivatives by prime, one gets
\begin{align}
\partial_y \langle uv\rangle+\alpha U&=0\,,
\label{eq:mom1}\\\!\!\!
\partial_y\left\langle \!v\left(p+\frac{ u^2+v^2 }{2}\right)\right\rangle&=\epsilon -U'\langle uv\rangle-\alpha \left\langle u^2+v^2 \right\rangle.\label{eq:energybalance}
\end{align}

To make analytical progress, one usually assumes that fluctuations are suppressed by the mean flow and employs the quasi-linear approximation, neglecting the cubic terms in Eq.~\eqref{eq:energybalance}. 
The assumption seems to be supported both by the energy argument (most of the energy is transferred from fluctuations to the mean flow so one expects
interactions  of  fluctuations to be unimportant) and by the momentum balance in Eq.~\eqref{eq:mom1} which gives $\langle uv\rangle\simeq\sqrt{\delta}(\epsilon L)^{2/3}$. For a circular vortex it was assumed additionally that the whole energy flux divergence is negligible~\cite{laurie_universal_2014}, including the pressure term.   Using this assumption, Eq.~\eqref{eq:energybalance} is reduced to $\epsilon=U'\langle uv\rangle$, resulting in a closed system for $\langle uv\rangle$ and $U$ as $\delta\to0$.  Jets, however, have lines with $U'=0$, where we expect the energy flux divergence to be comparable to $\epsilon$.

For simulations A and F, Figs.~\ref{fig:6}(a-b) show the mean flow profiles, Figs.~\ref{fig:6}(c-d) present the terms of~\eqref{eq:energybalance}. To reduce noise, the data were low-pass filtered using a Gaussian kernel in Fourier space with an effective cutoff $L/8$.
Under long-time (or zonal) averaging, the vortices contribute both to the mean flow and the fluctuations, making them strong.
The locations of the vortices are thus characterized by peaks in the energy dissipation of the fluctuations.

For the dipole, the energy flux divergence is important everywhere, and in particular the cubic terms are not small. There is no place where the quasi-linear approximation can be applied. In the three-vortex configuration, simulation F, the energy flux divergence is negligible in a small region between the same-sign vortices, the approximation of~\cite{laurie_universal_2014} seems to be applicable there. The black curve in Figs.~\ref{fig:6}(c-d), representing the sum of terms in \eqref{eq:energybalance} without $\epsilon$, shows convergence of the statistics. Naturally, it is worse in the region of vortices where fluctuations are stronger.

It is also illuminating to consider the balance of enstrophy (squared vorticity) for the fluctuations:
\begin{equation}
\frac{1}{2} \partial_y \left\langle v \omega^2\right\rangle=Q-\left\langle v\omega\right\rangle \Omega',
\end{equation}
where $\omega$ and $\Omega=-U'$ are the fluctuating and mean vorticity respectively, and $Q= \eta-\nu \left\langle\omega \left(-\nabla^2\right)^p \omega\right\rangle-\alpha \left\langle\omega^2\right\rangle$ with $\eta=\left\langle\omega \left(\nabla\times{\bf f}\right)\cdot\hat{z}\right\rangle$. Most of the injected enstrophy should be dissipated in the direct cascade by viscosity,  so that when integrated over $y$, $Q$ is a small but finite rate of enstrophy absorption by the mean flow.
The turbulence-flow enstrophy exchange term, $\left\langle v\omega\right\rangle \Omega'= \alpha U \Omega'$, turns into zero where $\Omega'=0$ and  $U=0$. That hints that the cubic term (turbulent flux divergence) may be large there and the quasi-linear approximation invalid.
Indeed, we find in Fig.~\ref{fig:6}(e-f) that 
the balance is everywhere dominated by $Q$ and the turbulent flux, which goes from the jets  to vortices, where viscous dissipation is larger. The quasi-linear approximation thus fails even in the regions where the velocity cubic terms are small in Fig.~\ref{fig:6}(c-d).

To conclude, let us reiterate the hierarchy of fluctuations and symmetries.
The cascade-related weak fluctuations with velocities $v_L\simeq(\epsilon L)^{1/3}$ average to zero on a timescale exceeding their correlation time $\tau_f=L/v_L$. Vortices and jets have much larger velocities $U\simeq \sqrt{\epsilon/\alpha}$   and persist in the steady state.
Fluctuations make the vortices wander along the jets, so averaging in a fixed reference frame over times exceeding $\tau_f$ (and hence exceeding their turnover time $L/U$) makes the flow unidirectional and restores the translation invariance along the short direction $x$.
One may expect that averaging over even longer timescales gives zero mean flow and thus restores translational invariance along both $x$ and $y$. However, we found such a time only for a square box where the vortex dipole wanders around because the distance between vortices fluctuates. In the rectangle, the longest averages
(over times exceeding $1/\alpha$) give stable jets and thus do not restore the translation invariance along $y$.
Apparently, it is much easier to move vortices or dipoles than it is to shift jets.
Note also the remarkable breakdown of reflection symmetry by the appearance of a third vortex at lower friction.
This work provides only a first glimpse into the intricacies of flows created by an inverse cascade in a box with a globally broken $x-y$ symmetry. In geophysical systems one also has differential rotation ($\beta$-effect), which breaks the symmetry locally and is expected to destroy the large-scale vortices. We leave the complete characterization of the parameter space $\delta$, $l_x$, $\beta$ for a future publication.

\begin{acknowledgments}
{\it \textbf{Acknowledgments.}} The authors are grateful to T.~Grafke, V.~Lebedev and P.~Wiegmann for illuminating discussions.
A.F. was supported by the Adams Fellowship Program of the Israel
Academy of Sciences and Humanities. The work of G.F. was supported by the grant 882 of the Israel Science Foundation and the project 14-22-00259 of the Russian Science Foundation.

A.F. and J.L. contributed equally to this work.
\end{acknowledgments}

\section*{Supplementary}	
\noindent{\it \textbf{Numerical parameters.}}
In total, we have performed eight numerical simulations of the two-dimensional Navier-Stokes equation, labeled A-H. We used a pseudo-spectral spatial discretization and a fourth-order exponential time-differences temporal scheme. Our simulations were done in a periodic rectangular box of size $2\pi l_x \times 2\pi$ with resolution $512l_x \times 512$ unless otherwise stated. As outlined in the main text, the stochastic forcing is defined in Fourier space in an annulus of $99 \leq k<101$ where there is an independent complex Gaussian white noise acting on each mode with amplitude $=0.1$. We characterize the flows by the small parameter $\delta = \epsilon^{-1/3}L^{2/3}\alpha$ which relates the strength of the turbulence to that of the mean flow. In all cases, we find that almost all of the enstrophy injected, $\eta_f$, is dissipated by the hyper-viscous term at high wavenumbers, whereas a significant proportion of the energy injected $\epsilon_f$ is also dissipated at high wavenumbers. Consequently, we estimate the inverse energy flux $\epsilon$ by computing the energy dissipation rate by friction during the steady state regime. The following tables detail the physical parameters used in all our simulations:

\begin{table}[htbp]
\begin{tabular}{| c | c | c | c |}
  \hline
  	 & $A$ & $B$ & $C$\\
  	\hline
  	$l_x$ & $1/2$ & $3/4$ & $1$\\
  \hline
  	$dt$ & $2\times 10^{-3}$ & $2 \times 10^{-3}$ & $2\times 10^{-3}$ \\
  \hline
   $\epsilon_f$ &   $3.09\times 10^{-4}$&  $4.69\times 10^{-4}$& $6.27\times 10^{-4}$ \\
  \hline
   $\eta_f$ &   $3.09$&  $4.69$& $6.26$ \\
  \hline
  $\epsilon$ &   $2.27\times 10^{-4}$&  $3.43\times 10^{-4}$& $4.67\times 10^{-4}$ \\
  \hline
  $\delta$ & $5.58\times 10^{-3}$ & $4.86\times 10^{-3}$ & $4.39\times 10^{-3}$\\
  \hline
  $\tau_m$ & $4.17$ & $3.39$ & $2.91$\\
  \hline
  $\tau_f$ & $55.82$& $48.64$ & $43.89$\\
  \hline
  $\tau_d$ & $1\times 10^{4}$ & $1.0\times 10^{4}$ & $1 \times 10^{4}$ \\
  \hline
\end{tabular}
\caption{For simulations A-C, we set the hyper-viscous and friction dissipation coefficients to be $\nu =1\times 10^{-36}$, $\alpha = 1\times 10^{-4}$. The energy and enstrophy injection rates differ due to the difference in resolution (number of Fourier wavenumbers in the forcing annulus).}
\end{table}

\begin{table}[htbp]
\begin{tabular}{| c | c | c | c |}
  \hline
  	& $D$ & $E$ & $F$\\
  \hline
  	$dt$ & $2.5\times 10^{-3}$ & $1\times 10^{-3}$ & $5\times 10^{-4}$ \\
  	\hline
	$\alpha$ & $2\times 10^{-4}$  & $5\times 10^{-5}$ & $3\times 10^{-5}$ \\
  \hline
  $\epsilon$ &   $2.28\times 10^{-4}$&  $2.30\times 10^{-4}$& $2.30\times 10^{-4}$ \\
  \hline
  $\delta$ & $1.12\times 10^{-2}$ & $2.78\times 10^{-3}$ & $1.67\times 10^{-3}$\\
  \hline
  $\tau_m$ & $5.89$ & $2.93$ & $2.27$\\
  \hline
  $\tau_f$ & $55.74$& $55.57$ & $55.57$\\
  \hline
  $\tau_d$ & $5\times 10^{3}$ & $2\times 10^{4}$ & $3.33 \times 10^{4}$ \\
  \hline
\end{tabular}
\caption{Simulations D-F are all performed in a periodic box with aspect ratio $l_x=1/2$, with $\nu =1\times 10^{-36}$. Resultantly, the energy and enstrophy injection rates are $\epsilon_f = 3.09\times 10^{-4}$ and $\eta_f = 3.09$ respectively.}
\end{table}

\begin{table}[htbp]
\begin{tabular}{| c | c | c |}
  \hline
  	& $G$ & $H$\\
  	\hline
  	$l_x$ & $1/3$ & $1/4$\\
  \hline
  	$dt$ & $2\times 10^{-3}$ & $2\times 10^{-3}$\\
  \hline
   $\epsilon_f$ & $2.26\times 10^{-4}$&  $1.67\times 10^{-4}$\\
  \hline
   $\eta_f$ & $2.25$&  $1.67$\\
  \hline
  $\epsilon$ & $1.64\times 10^{-4}$&  $1.24\times 10^{-4}$\\
  \hline
  $\delta$ & $6.22\times 10^{-3}$ & $6.83\times 10^{-3}$ \\
  \hline
  $\tau_m$ & $4.91$ & $5.64$\\
  \hline
  $\tau_f$ & $62.21$& $68.28$\\
  \hline
  $\tau_d$ & $1\times 10^{4}$ & $1\times 10^{4}$ \\
  \hline
\end{tabular}
\caption{Simulations G and H are performed in a periodic box with resolutions $128\times 384$ and $128\times 512$ respectively. Both simulations used $\nu =1\times 10^{-36}$ and $\alpha = 1\times 10^{-4}$.}
\end{table}

\begin{figure}[!ht]
\begin{center}
\includegraphics[width=0.8\columnwidth]{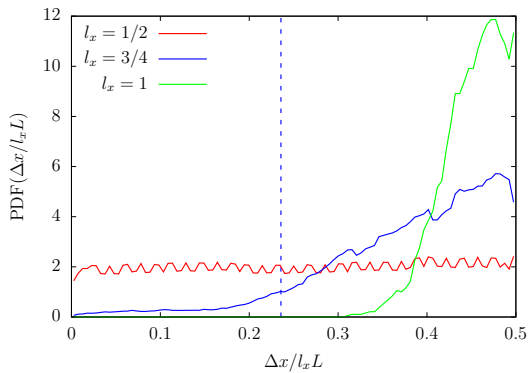}
\caption{(color online) PDF of relative $x$-coordinate separation, scaled by the box length in $\hat{x}$, of the two vortices centers in simulations A (red), B (blue), and C (green). The vertical blue dashed line indicates the minimum distance prediction for $l_x=3/4$.   \label{fig:app1}}
\end{center}
\end{figure}

{\it \textbf{Vortex motion and timescales.}}
In the main text (see Fig.~3) we comment on the relative movement of the two vortices and make a prediction
that the critical aspect ratio $l_x$ below which the second vortex can move across the entire short side of the rectangular periodic box is close to $1/\sqrt{2}$.  In Fig.~\ref{fig:app1}, we quantify this movement by plotting the PDF of the separation between the two vortices centres in the horizontal $x$-direction for simulations A-C. We observe that for $l_x=1/2$, in the frame of the first vortex, there is almost uniform probability of observing the second vortex anywhere across the box (red curve). This is to be expected assuming that our minimal dipole separation prediction is correct. Aspect ratios $l_x=3/4$ and $1$ are larger than our critical value, therefore, we expect that the horizontal movement of the second vortex will be limited. For the square box (green curve), the PDF is peaked around one half indicating little or no fluctuations from the rigid diagonal lattice. For aspect ratio $l_x=3/4$, we observed that the vortex separation is mainly contained at large distances (blue curve). The vertical blue dashed line indicates our prediction of the minimal separation distance between the two vortices.  In both cases, $l_x=3/4$ and $1$, slight deviations are observed due to the presence of turbulent fluctuations. Such fluctuations cause deviations in all directions, and in particular in the $y$-direction, that is slight relative vertical motion does exist for a rectangular box. In summary, the evidence is compatible with our prediction that $l_x = 1/\sqrt{2}$ is a critical aspect ratio below which dynamical restrictions on the motion of the second vortex are absent.

\begin{figure}[!ht]
\begin{center}
\includegraphics[width=\columnwidth]{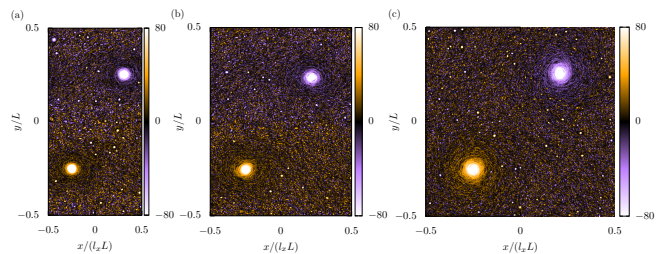}
\caption{(color online) Scaled total vorticity $\sqrt{\alpha L^2/\epsilon}\left(\nabla \times {\bf v}\right)$ during the steady state regime for simulations A (a)
, B (b), C (c) take at a single snapshot.\label{fig:app2}}
\end{center}
\end{figure}

\begin{figure}[!ht]
\begin{center}
\includegraphics[width=\columnwidth]{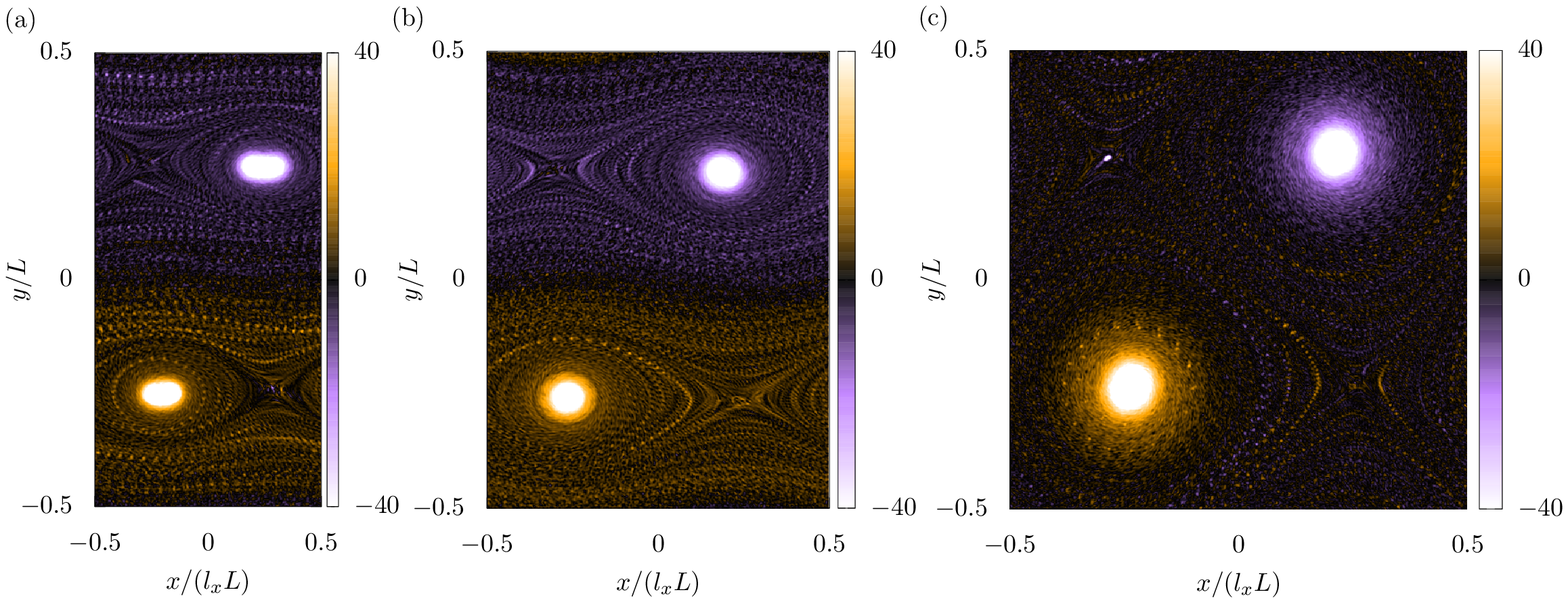}
\caption{(color online) Scaled total vorticity $\sqrt{\alpha L^2/\epsilon}\left(\nabla \times {\bf v}\right)$ during the steady state regime for simulations A (a)
, B (b), C (c) time averaged over $\tau_m\equiv\sqrt{\alpha L^2/\epsilon}$. \label{fig:app3}}
\end{center}
\end{figure}

\begin{figure}[!ht]
\begin{center}
\includegraphics[width=\columnwidth]{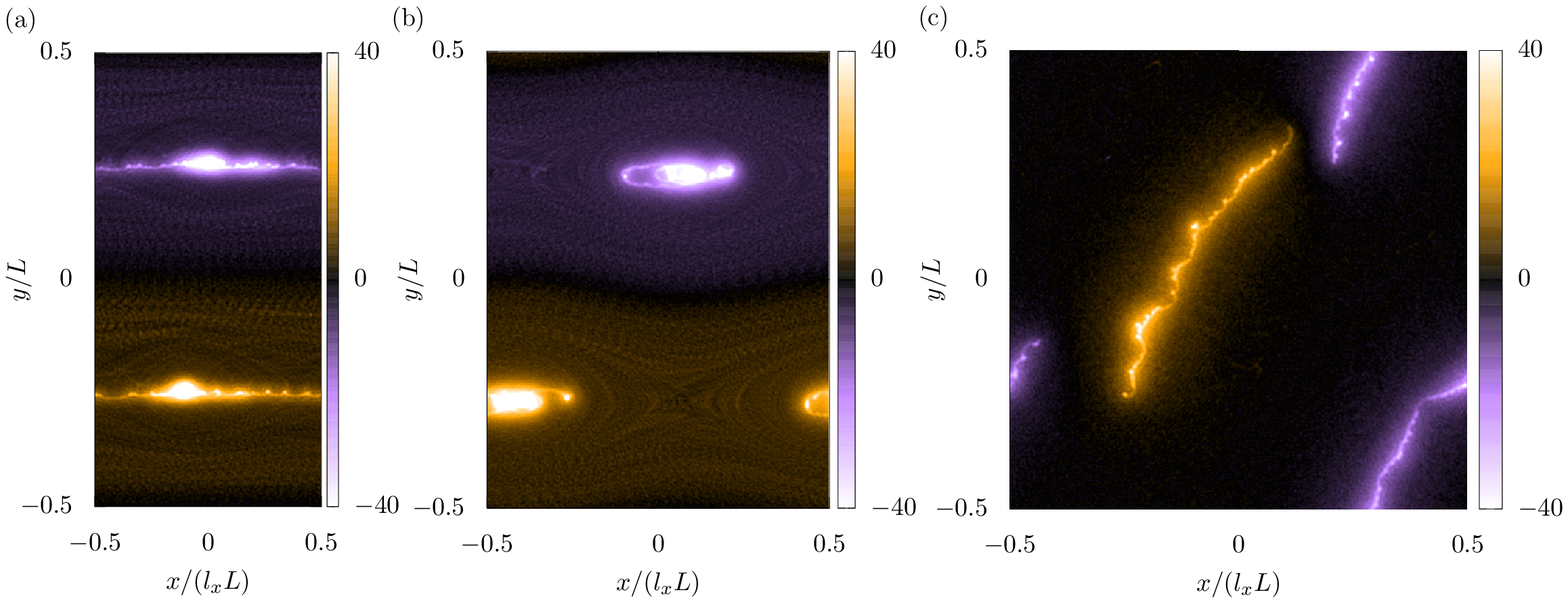}
\caption{(color online) Scaled total vorticity $\sqrt{\alpha L^2/\epsilon}\left(\nabla \times {\bf v}\right)$ during the steady state regime for simulations A (a)
, B (b), C (c) time averaged over $\tau_f \equiv \epsilon^{-1/3}L^{2/3}$.   \label{fig:app4}}
\end{center}
\end{figure}

\begin{figure}[!ht]
\begin{center}
\includegraphics[width=\columnwidth]{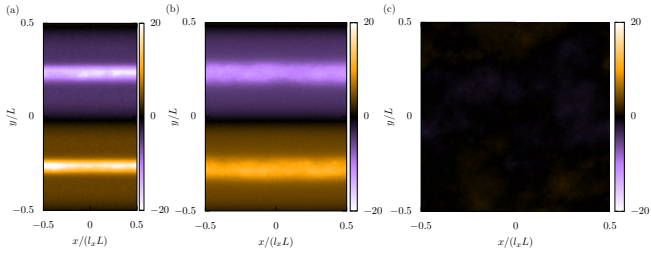}
\caption{(color online) Scaled total vorticity $\sqrt{\alpha L^2/\epsilon}\left(\nabla \times {\bf v}\right)$ during the steady state regime for simulations A (a)
, B (b), C (c) time averaged over $\tau_d\equiv 1/\alpha$.  \label{fig:app5}}
\end{center}
\end{figure}

In Figs.~\ref{fig:app2}-\ref{fig:app5} we plot the scaled vorticity of simulations A-C for a snapshot and also for temporal averages over times $\tau_m$, $\tau_f$, and $\tau_d$ respectively.  In all cases, the averaging is performed in the fixed reference frame not following vortices and jets. What we observe in Fig.~\ref{fig:app3} is that averaging over the mean flow timescale $\tau_m$ results in averaging out the fluctuations and clear appearance of the jets in the non-square boxes. The vortices are almost static during such time. Indeed, the mean vorticity level set is similar to the streamline plot of Fig.~2 in the main text implying that the mean flow on such timescales is a solution of the Euler equation, satisfying  $\langle {\bf v}\rangle \cdot \nabla \Omega=0$. Fig.~\ref{fig:app4} shows averaging over the fluctuation timescale $\tau_f$, where we observe some motion of the vortices. For aspect ratio $l_x=1/2$, Fig.~\ref{fig:app4}(a), we see the spreading of the vortices 
and formation of two strong jets associated to the zonal drift of the vortices. Interestingly, for $l_x=3/4$, Fig.~\ref{fig:app4}(b), the vortices are still relatively static, although we do observe some preliminary spreading of the vortices along the $x$-direction. For the square domain, the vortex dipole begins to wander through the domain as a coherent object.  There is no preferential direction, and so we expect this wandering to be randomly orientated, dictated by the fluctuations in the relative position of the vortices. Such fluctuations are correlated on the timescale $\tau_f$ so that the dipole motion observed in Fig.~\ref{fig:app4}(c) takes the form of almost straight lines.
Finally, in Fig.~\ref{fig:app5}, we observe the vorticity field after averaging over the longest dynamical timescale $\tau_d$ associated to the time for large-scale dissipation to have an effect.  Surprisingly, we still observe two coherent jets in the panels (a) and (b).  This implies that fluctuations in the vertical direction are 
suppressed by the generation of the zonal jets. On the other hand, in the square domain, shown in the panel (c), we observe almost no mean flow because the vortex dipole randomly wanders around the whole domain averaging to zero, which over the long timescales of order $\tau_d$.

\begin{figure}[!ht]
\begin{center}
\includegraphics[width=\columnwidth]{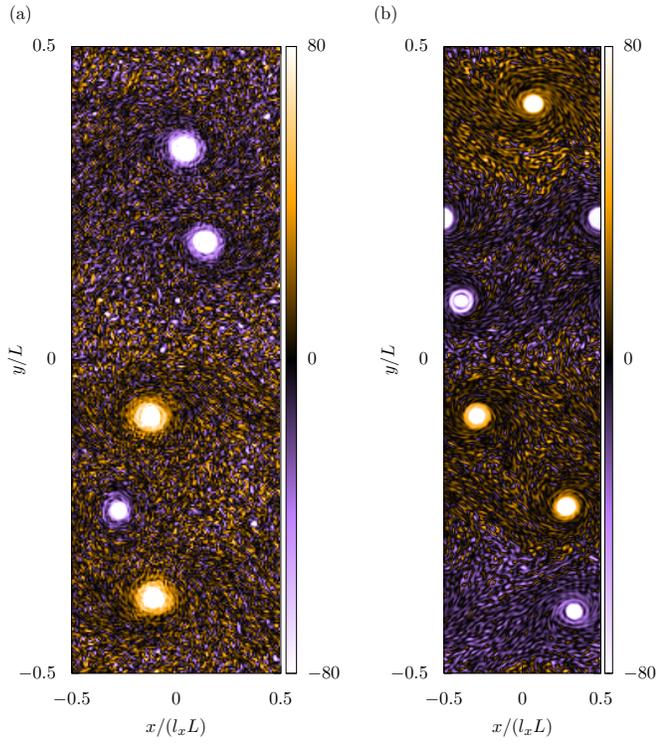}
\caption{(color online) Plots of the scaled total vorticity $\sqrt{\alpha L^2/\epsilon}\left(\nabla \times {\bf v}\right)$ during the statistical steady state regime for simulations G (a) and H (b).   \label{fig:app6}}
\end{center}
\end{figure}

{\it \textbf{Jets and vortices in large aspect ratios.}}
We discussed in the main text that we performed additional simulations, G and H, in larger aspect ratios, namely $l_x = 1/3$ and $1/4$.  In Fig.~\ref{fig:app6}, we present the vorticity snapshot during steady state conditions of these simulations. We observe the appearance of multiple vortices and jets. For aspect ratio $l_x=1/3$, we see two zonal jets, similar to what was observed in simulations A, B, and D-F, but with additional vortices, in this case five. We do also obverse switches to a regime with four (two positive plus two negative) vortices. In Fig.~\ref{fig:app6}(b), which corresponds to aspect ratio $l_x=1/4$, we observe four jets and six vortices. In summary, we observe a most surprising feature: namely that the number of jets is not conserved as the aspect ratio is increased. This gives a strong indication that the gravest mode argument is wrong. Finally, it should be noted that in both cases, $l_x=1/3$ and $1/4$, the number of vortices present in Fig.~\ref{fig:app6} do not violate our prediction for the minimal distance between vortices, $l_xL/\sqrt{2}$.

\end{document}